\documentclass{article}

\usepackage{framed}
\usepackage{amssymb,amsfonts,amsmath,mathtext,cite,enumerate,float,amsthm}
\usepackage{ytableau}
\usepackage[vcentermath]{youngtab}
\usepackage{tikz}
\usetikzlibrary{positioning}
\usetikzlibrary{arrows}
\graphicspath{ {C:/Users/YourFrienD/Documents/} }

\hoffset= - 1.0in

\newtheorem{statement}{Statement}

\newcommand{\eqnb}{\begin{equation}}
\newcommand{\eqn}{\end{equation}}

\newcommand{\h}{\hbar}

\newcommand{\Al}{\mathcal{A}l}
\newcommand{\K}{\mathcal{K}}

\newcommand{\A}{\mathcal{A}}
\newcommand{\fullKP}{\widehat{\mathcal{KP}}}
\newcommand{\KP}{\mathcal{KP}}

\begin{document}

\title{\vspace{0.1cm}{\Large {\bf Colored Alexander polynomials and KP hierarchy}\vspace{.2cm}}
\author{
{\bf A. Mironov$^{a,b,c}$}\footnote{mironov@lpi.ru; mironov@itep.ru},
{\bf S. Mironov$^{b,d}$}\footnote{sa.mironov\_1@physics.msu.ru},
{\bf V. Mishnyakov $^{b,e}$}\thanks{mishnyakovvv@gmail.com},
{\bf A. Morozov$^{b,c}$}\thanks{morozov@itep.ru},
{\bf A. Sleptsov$^{b,c,f}$}\thanks{sleptsov@itep.ru}} \date{ }%
}
\maketitle

\vspace{-5cm}

\begin{center}
	\hfill FIAN/TD-07/18\\
	\hfill IITP/TH-09/18\\
	\hfill ITEP/TH-11/18
\end{center}

\vspace{2.3cm}

\begin{center}

$^a$ {\small {\it Lebedev Physics Institute, Moscow 119991, Russia}}\\
$^b$ {\small {\it ITEP, Moscow 117218, Russia}}\\
$^c$ {\small {\it Institute for Information Transmission Problems, Moscow 127994, Russia}}\\
$^d$ {\small {\it Institute of Nuclear Research, Moscow 117312, Russia}}\\
$^e$ {\small {\it Moscow State University, Physical Department,
Vorobjevy Gory, Moscow, 119899, Russia }}\\
$^f$ {\small {\it Laboratory of Quantum Topology, Chelyabinsk State University, Chelyabinsk 454001, Russia }}

\end{center}

\vspace{1cm}

\begin{abstract}

We discuss the relation between knot polynomials and the KP hierarchy.
Mainly, we study the scaling 1-hook property of the coloured Alexander polynomial: $\A^\K_R(q)=\A^\K_{[1]}(q^{\vert R\vert})$ for all 1-hook Young diagrams $R$.
Via  the Kontsevich construction, it is reformulated as a system of linear equations.
It appears that the solutions of this system induce the KP equations in the Hirota form.
The Alexander polynomial is a specialization of the HOMFLY polynomial, and it is a kind of a dual to the double scaling limit, which gives the special polynomial,
in the sense that, while the special polynomials provide solutions to the KP hierarchy,
the Alexander polynomials provide the equations of this hierarchy.
This gives a new connection with integrable properties of knot polynomials
and puts an interesting question about the way the KP hierarchy is encoded
in the full HOMFLY polynomial.
\end{abstract}

\vspace{.5cm}



\section{Introduction}

Nowadays knot theory is of great interest in mathematical physics.
This is due to the fact that knot invariants appear in various
physical problems such as quantum field theories \cite{Witten,ChernSimons,CS}, quantum groups \cite{RT}, lattice models \cite{StatPhys}, CFT \cite{WZNW}, topological strings \cite{TopStrings},
quantum computing \cite{MMMMM} etc.
These correspondences lead to generalizations of some already known invariants
and to discoveries of new ones.

The class of polynomial invariants is probably the most developed and the most actively studied.
One of the most important in this class is the (unreduced) coloured HOMFLY polynomial ${\cal H}_R^\K(q,a)$ of the knot ${\cal K}$ coloured with representation $R$.
It takes values in the ring of Laurent polynomials of two variables
  $\mathbb{Z}[q,q^{-1},a,a^{-1}]$.
It may be
defined as the vacuum expectation value of a Wilson loop along the knot
in Chern-Simons gauge theory,
with the gauge group $G=SU(N)$ and representation $R$ \cite{Witten,Hidden}:
\begin{equation}
{\cal H}_R^\K(q,a)=\dfrac{1}{Z}\int DA \ \  e^{-\frac{i}{\h}S_{CS}[A]} \ W_R(K,A),
\end{equation}
where the Wilson loop and the Chern-Simons action given by
\begin{equation}
W_R(K,A)  =  \mathrm{ tr_R } \ P \mathrm{exp} \left(\oint A_\mu^a (x) T^a dx^\mu \ \right) , \qquad S_{CS}[A]={\kappa\over 4\pi}\int_M \mathrm{Tr}(A \wedge dA +\dfrac{2}{3} A\wedge A \wedge A).
\end{equation}
The variables $q, a$ in the HOMFLY polynomial are
$$q=e^\h, \ a=e^{N \h},\ \ \ \h:={2\pi i\over \kappa+N}.$$
In such a parametrization, the polynomial may be represented as a series in the variable  $\h$.
Furthermore, there is a natural "quasiclassical" double scaling expansion given by
$
\h \rightarrow 0 , N \rightarrow \infty$ such that $N\h$ stays fixed.
In other words, this means taking $q=1$ and keeping the variable $a$ arbitrary.
The polynomials that emerge as the special value of the HOMFLY polynomials ${\cal H}^\K_R(1,a)=\sigma_R^\K(a)$ are called the special polynomials \cite{DMMSS}.
Their $R$ dependence has a simple power-like form,
which expresses them through the special polynomial in the fundamental representation \cite{DMMSS,Itoyama,Zhu,GenusExpansion}:
\begin{equation}
\sigma_R^\K(a)=\left(\sigma_{[1]}^\K(a) \right)^{|R|}.
\end{equation}
Hereafter, we identify the representation $R$ with the Young diagram associated with it: $R=\{R_i\},\ R_1\ge R_2\ge\ldots\ge R_{l(R)}$,
$|R|:=\sum_i R_i$.

These polynomials exhibit integrable properties \cite{GenusHurwitz,GenusExpansion}:
their $R$ dependence allows one to construct from them a KP $\tau$-function, which is the value of the
Ooguri-Vafa partition function:
\begin{equation}
\mathcal{Z}^{\K}(\bar{p}|a,q) =  \sum_R {\cal H}^{\K}_R(a,q)D_R\chi_R\{\bar{p}\},
\end{equation}
at $q=1$. In this formula, $\chi_R\{p\}$ is the Schur polynomial in the representation $R$, $D_R=\chi_R\{p^*\}$ with $p_k^*=\frac{a^k-a^{-k}}{q^k-q^{-k}}$ is the quantum dimension.

On the other hand, the ``dual" limit would be taking $a \rightarrow 1$ and leaving us with ${\cal H}_R^\K(q,1)$. For  the fundamental representation $R$, this gives the oldest polynomial knot invariant, the Alexander polynomial.
Its coloured version displays a ``dual" property  with respect to $R$ \cite{Itoyama,MM}\footnote{Let us note that our notion of coloured Alexander polynomial is totally different from that defined in \cite{Mur}.}:
\begin{equation}\label{alexproperty0}
\A^\K_R(q)=\A^\K_{[1]}(q^{\vert R\vert})\ ,\quad \text{where} \ R=[r,1^L].
\end{equation}
which holds only for the representations corresponding to 1-hook Young diagrams. We have studied this property perturbatively and discovered that it also miraculously related to the KP hierarchy. \textbf{We found that, while the special polynomials provide solutions to the KP hierarchy, the Alexander polynomials induce the equations of the KP hierarchy}. In this paper, we state this result giving the shorter half of the proof. Explicit calculations and the detailed proof will be presented elsewhere.\\
This observation not only gives another example of integrable properties of knot polynomials, but also argues in favour of use of the term ``dual" in discussing the two limits of the HOMFLY polynomial. The results can be summarized in the following diagram: \\




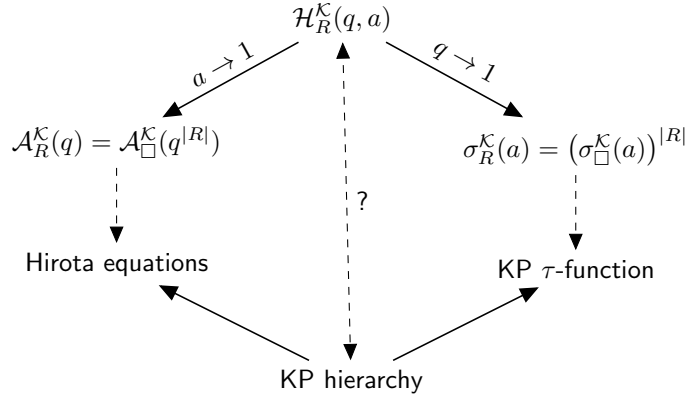
\begin{center}
\begin{tikzpicture}[>=triangle 45,font=\sffamily]
    \node (H) at (0,0) {${\cal H}_R^\K(q,a)$};
    \node (Y) [below left=1cm and 0.7cm of H]  {${\cal A}^\K_R(q)={\cal A}^\K_{\square}(q^{|R|})$};
    \node (Z) [below right=1cm and 0.7cm of H] {$\sigma^\K_R(a)=\left(\sigma^\K_{\square}(a)\right)^{|R|}$};
    \node (T1) [below =1cm of Z]  {KP $\tau$-function};
    \node (T2) [below =1cm of Y]  {Hirota equations};
    \node (T3) [below right=1cm and 0.7cm of T2]  {KP hierarchy};
    \draw [semithick,->] (H) -- (Y) node [midway,above,sloped] {$a\rightarrow 1$};
    \draw [semithick,->] (H) -- (Z)node [midway,above,sloped] {$q\rightarrow 1$};
    \draw [dashed,->] (Y) -- (T2);
    \draw [dashed,->] (Z) -- (T1);
    \draw [semithick, ->] (T3) -- (T2);
    \draw [semithick, ->] (T3) -- (T1);
    \draw [dashed, <->] (T3) -- (H) node [midway,right] {?};
\end{tikzpicture}
\end{center}

Another well-known perturbative expansion of the HOMFLY polynomial is the loop expansion \cite{Hidden}, which is based on the gauge invariance of Chern-Simons theory. Evaluating the Wilson loop correlators can be done in some fixed gauge. In the temporal gauge \cite{Sm}, $A_0=0$, the Wilson loop acquires the polynomial form of the coloured HOMFLY invariant.
When calculated in the holomorphic gauge \cite{La} $A_x+iA_y=0$, it gives the Kontsevich integral \cite{ChmutovDuzhin,DBSlSm}. The theory is gauge invariant, therefore the two object are equal, however, the Kontsevich integral is a perturbative expansion and the HOMFLY polynomial is not. Therefore this construction gives a perturbative description of the HOMFLY polynomial with arbitrary variables $q,a$. It appears to have a nicely looking structure \cite{Labast}:
\begin{equation}\label{Expansion}
{\cal H}_R^\K=\sum_n \left(\sum_j v_{n,j}^\mathcal{K} r_{n,j}^R\right) \h^n.
\end{equation}
A remarkable fact is that the knot dependence and the group theoretic dependence split explicitly. The group one is represented by the so called group factors $r_{n,j}^R$. They are group invariants that appear in the Kontsevich integral as some trivalent diagrams, which are further expressed as traces of products of the algebra generators $T^a$ and the structure constants:
\begin{equation}
r^R_{n,j} \sim \mathrm{tr}_R(T^{a_1}\ldots T^{a_n}).
\end{equation}
\\
The knot dependent part is $v_{n,j}^\mathcal{K}$, which are some numerical invariants. Another point is that they appear to be exactly the famous Vassiliev invariants or invariants of finite type\cite{Labast,CDIntro}. These numerical invariant are considered as potential candidates for a complete set of invariant and are therefore important to study. In this paper, however, we mostly focus on the group theoretic part.

In section 2, we describe the Alexander polynomial and its basic property, in section 3, we discuss a set of equations that originate from this basic property, and, in section 4, we review the Hirota bilinear identities and the KP hierarchy in order to compare them, in section 5, with the Alexander set of equations. In section 5, we demonstrate that the Hirota KP bilinear equations are satisfied when the Hirota derivatives are replaced by the Casimir eigenvalues in the 1-hook diagrams, and this solution is equivalent to the solution of the Alexander set of equations. This is our main result in this paper, while another our result is the number of actually different KP equations of each order which coincides with the number of solutions of the Alexander equations.

\section{Alexander polynomial}
The Alexander polynomial is a knot invariant in the ring of Laurent polynomials in one variable $\mathbb{Z}[q^{-1},q]$. Originally, it is defined via the $H_1(X_{\infty})$ homology group of the infinite cyclic cover of the knot complement $S^3 \setminus \K$ and is denoted as $\A(q)$ \cite{Prasolov}.\\

Apart from its pure topological construction, it also appears as a specific value of the ordinary (fundamental) HOMFLY polynomial \cite{CDIntro}:
\begin{equation}
\A^\K(q)={\cal H}^\K(q,1).
\end{equation}
As higher representations of the gauge group lead to the coloured HOMFLY polynomials, one can immediately define the coloured Alexander polynomials
\begin{equation}\A^\K_R(q)={\cal H}^\K_R(q,1) \quad \text{or} \quad\ \A^\K_R(e^\h)=\lim_{N \rightarrow 0} {\cal H}^\K_R(e^\h,e^{N\h}).
\end{equation}
As we already mentioned \eqref{alexproperty0}, this polynomial has a peculiar $R$ dependence, but only for special representations.

A 1-hook Young diagram is a diagram of the form $\lambda=[r,\underbrace{1,\ldots,1}_L]$:
\begin{gather*}
\yng(5,3,2,1) \ \ \ \text{[5,3,2,1] diagram,}\qquad \
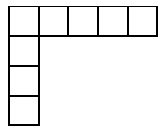
\yng(5,1,1,1) \ \ \ \text{1-hook diagram [5,1,1,1]}
\end{gather*}
 \\
Now, for any knot $\K$ and for any 1-hook Young diagram $R=[r,1^L]$, \cite{Itoyama,MM}
\begin{equation}\label{a-a}
\A^\K_R(q)=\A^\K_{[1]}(q^{\vert R\vert})\ ,\quad \text{where} \ |R|=r+L.
\end{equation}

Another important property is the symmetry of the HOMFLY (and, in particular, Alexander) polynomials with respect to the transposition of the Young diagram of the representation. This property holds for arbitrary diagrams $R$ and comes from the corresponding property of quantum groups
and WZW theories \cite{RankLevel1}\cite{RankLevel2}, in the latter case, it is called the rank-level duality (see also \cite{LiuPeng}):
\begin{equation}
{\cal H}_R^\K(q,a)={\cal H}_{R^T}^\K(q^{-1},a).
\end{equation}
This property is immediately inherited by the Alexander polynomials,
\begin{equation}\label{ranklevel}
\A^\K_R(q)=\A^\K_{R^T} (q^{-1}).
\end{equation}

\section{Alexander system of equations}

Let us now concentrate on the basic property (\ref{a-a}) of the Alexander polynomial and look at it perturbatively using the expansion (\ref{Expansion}). As a specialization of the HOMFLY invariant, the Alexander polynomial inherits its general structure.
Let us substitute the expansion \eqref{Expansion} in  \eqref{a-a}, put $N=0$ and denote the resulting group factors as $A_{i,j}$:
\begin{equation}
\A^\K_R(q)-\A^\K_{[1]}(q^{\vert R\vert})=\sum_n \h^n \sum_m {v}^\K_{n,m}
\left(r^R_{n,m}-|R|\cdot r^{[1]}_{n,m}\right)\Big|_{N=0}=:\sum_n \h^n \sum_m {v}^\K_{n,m} A^R_{n,m}\ \stackrel{^{R=[r,1^L]}}{=}\ 0.
\end{equation}
This equality should hold at all orders of $\h$. Moreover, since the Vassiliev invariants depend on the knot, one arrives at the property of the Alexander group factors
\begin{equation}\label{16}
\boxed{
A^{[r,1^L]}_{n,m}= 0.
}
\end{equation}
The group factors $r^R_{n,j}$, being group invariants, can be expanded into the basis of the Casimir invariants of the algebra \cite{Labast}:
\begin{equation}\label{16p}
A^R_{n,m} = \sum_{|\Delta|\le n} \alpha_{\Delta,m} C_\Delta(R),
\end{equation}
where we label the monomials of $C_k$ by the Young diagrams in accordance with
\begin{equation}
C_\Delta=\prod_{i=1}^{l(\Delta)} C_{\Delta_i}.
\end{equation}
Then, $A^R_{n,m}$ can be considered as functions of Casimir invariants only, all the dependence on the representation entering through these latter ones:
\begin{equation}
A^R_{n,m}=A_{n,m}(C)
\end{equation}
Note that one can also re-expand the difference
\begin{eqnarray}\label{17}
\A^\K_R(q)-\A^\K_{[1]}(q^{\vert R\vert})=\sum_n \h^n \sum_{|\Delta|\le n} C_\Delta(R) \sum_m {v}^\K_{n,m}  \alpha_{\Delta,m}=
\sum_n \h^n \sum_{|\Delta|\le n} \alpha^\K_\Delta C_\Delta(R),\\
\alpha^\K_\Delta:=\sum_m {v}^\K_{n,m}  \alpha_{\Delta,m}=(v_n^\K, \alpha_\Delta)\nonumber
\end{eqnarray}
into the Casimir invariants instead of the group factors.

Now let us consider general solutions to Eqs.(\ref{16})-(\ref{16p}) at a given level $n$:
\begin{equation}\label{gen}
X_{n,m} (C):= \sum_{|\Delta|=n} \xi_{\Delta}^{(m)} C_\Delta(R)=0
\end{equation}
for any 1-hook diagram $R$, where $\xi_\Delta$ are the coefficients in question, while the index $m$ labels the independent solutions of this equation. Equation \eqref{gen} should hold for each $L,r$ of the 1-hook Young diagram $[r,1^L]$, therefore, at each order, we get a system of equations on $\xi_\Delta$, which we call Alexander equations. The group-factors $A_{m,n}$ of the Alexander polynomial  are linear combination of the basis solutions to these equations:
\begin{equation}
A_{n,m}(C)\in \hbox{Span}\Big(\oplus_{k\le n}X_{k,m}(C)\Big)
\end{equation}
Therefore we consider (\ref{gen}) as equations defining the general structure of the polynomial.
Let us illustrate how it works. Let us choose the Casimir invariants as functions of the Young diagram $R$ in the following form \cite{Zhe}:
\begin{equation}
\label{cas}
C_k(R)=\sum_{i=1}\Big[(R_i-i+1/2)^k-(-i+1/2)^k\Big].
\end{equation}
Restricted to 1-hook diagrams $R=[r,\underbrace{1,\dots,1}_L]$, this formula reduces to:
\begin{equation}\label{casrl}
C_k(R)=(r-1/2)^k-(-L-1/2)^k,
\end{equation}
or, in a more symmetric form, with $l=L+1$ being the length of partition:
\begin{equation}\label{cas1}
C_k(R)=(r-1/2)^k+(-1)^{k+1}(l-1/2)^k.
\end{equation}
As a corollary of this explicit expression, the symmetry with respect to transposition of the diagram reads
\begin{equation}
C_\Delta(R^T)=(-1)^{|\Delta|+l(\Delta)}C_\Delta(R).
\end{equation}
It then follows that the solutions to the Alexander system of equations contain either only even number (denote them $X^e_{n,m}$), or only odd number of Casimir invariants in monomials (denote them $X^o_{n,m}$), i.e. either only even, or only odd monomials.

Some examples of solutions to (\ref{gen}) are then
\begin{align}\label{alexsol}
&\h^4:X^e_{4,1}(C)=C_1^4-4C_1 C_3+3 C_2^2;  \nonumber\\[4pt]
&\h^5: X^e_{5,1}(C)=C_2 C_1^3-3 C_4 C_1+2 C_2 C_3, \hspace{2.2cm} X^o_{5,1}(C)=C_1 \left( C_1^{4}-4\,C_1C_3+3\,C_2^{2} \right) ;\nonumber\\[4pt]
 &\h^6: X^e_{6,1}(C)=4 C_1^2 \left(C_1^4-4 C_3 C_1+3 C_2^2\right), \hspace{1.9cm} X^o_{6,1}(C)=C_1^{2}C_4-2\,C_1C_2C_3+C_2^{3} ,\nonumber \\
& \phantom{\h^6: } \ X^e_{6,2}(C)=2 C_3 C_1^3-3 C_2^2 C_1^2-8 C_3^2+9 C_2 C_4, \hspace{0.9cm} X^o_{6,2}(C)=C_2\left( C_1^{4}-4\,C_1C_3+3\,C_2^{2} \right),\nonumber\\
    &\phantom{\h^6: } \  X^e_{6,3}(C)=C_3 C_1^3+3 C_2^2 C_1^2-9 C_5 C_1+5 C_3^2;\\[4pt]
      &\h^7: X^e_{7,1}(C)=C_1 C_2 \left(C_1^4-4 C_3 C_1+3 C_2^2\right), \hspace{1.7cm} X^o_{7,1}(C)= C_1^{3}C_2^{2}-3\,C_1^{2}C_5+ 3\,C_1	C_3^{2}-C_2^{2}C_3, \nonumber\\
& \phantom{\h^7: } \ X^e_{7,2}(C)=C_1 \left(C_2 C_1^4-2 C_4
   C_1^2+C_2^3\right), \hspace{1.8cm} X^o_{7,2}(C)=C_1^{7}+9\,C_1^{2}C_5-25\,C_1 C_3^{2}+ 15\,C_2^{2}C_3,\nonumber\\
    &\phantom{\h^7: } \  X^e_{7,3}(C)=C_2 C_1^5-5 C_2^3 C_1-20 C_3 C_4+24 C_2 C_5, \quad X^o_{7,3}(C)=C_3 \left(C_1^{4}-4\,C_1C_3+3\,C_2^{2} \right), \nonumber\\
    &\phantom{\h^7: } \  X^e_{7,4}(C)=-C_2 C_1^5-3
   C_2^3 C_1+8 C_6 C_1-4 C_3 C_4, \hspace{0.4cm} X^o_{7,4}(C)=C_1^{2}C_5-C_1C_2C_4- C_1C_3^{2
   }+ C_2^{2}C_3;\nonumber
\end{align}
At the same time,
\begin{equation}
\A^\K_R(q)-\A^\K_{[1]}(q^{\vert R\vert})=\hbar^4 {v}^\K_{4,1} A_{4,1}(C)+\hbar^6\Big({v}^\K_{6,1} A_{6,1}(C)+{v}^\K_{6,2} A_{6,2}(C)+{v}^\K_{6,3} A_{6,3}(C)\Big)+\hbar^7{v}^\K_{7,1} A_{7,1}(C)+O(\hbar^8)
\end{equation}
with
\begin{eqnarray}
&A_{4,1}(C)=X^e_{4,1}(C) \nonumber\\
&A_{5,m}(C)=0\nonumber\\
&A_{6,1}(C)= X^e_{4,1}(C), \ \
A_{6,2}(C)={1\over 4}X^e_{6,1}(C), \ \  A_{6,3}(C)=-X^e_{6,1}(C) - \frac{5}{3}X^e_{6,2}(C) - \frac{2}{3}X^e_{6,3}(C) \\
&A_{7,1}(C)=X^o_{6,1}(C)
\end{eqnarray}
One can see that it may happen that not all of the solutions appear in the expansion itself: in this example, at the level 6, only two of three possible independent combinations emerge. In fact, the exact form is determined by other properties of the knot polynomial. For instance, by the general symmetry properties, there is no non-vanishing $A_{n,m}(C)$ at $n=5$, see also a more general statement \ref{s2} below. Generically, the number of $A_{n,m}(C)$ is less than the number of $X_{n,m}(C)$ at the given level $n$. Their number is an issue for future research.

Now let us discuss some other properties of the Alexander equations (\ref{gen}).


It is clear that the Casimir invariants can be chosen in infinitely many forms. Our particular choice is distinguished by the following two reasons:
\begin{enumerate}
\item with the choice (\ref{cas}) of the Casimir invariants, the corresponding Hurwitz partition function \cite{MMN1} becomes a KP tau-function  \cite{Casimir};
\item in terms of the Hurwitz partition function, this choice corresponds to the completed cycles and establishes a correspondence with the Gromov-Witten theory \cite{OP}.
\end{enumerate}


We are interested in the dimension of the space of solutions at each order of $\h$, which we denote as $\Al_n$. Moreover, we are interested in the dimension of the space of odd and even solutions (denoted $\Al^o_n \ , \ \Al^e_n$ respectively) separately. The dimension of the whole space if the sum of the latter. Now we simply state the result.
\begin{framed}
\begin{statement}
The dimensions of the vector spaces $\Al_n$ are expressed by the following formulas:
\begin{equation}\label{28}
\dim{\mathcal{A}l^e_n}=p_e(n)-\left[\dfrac{n}{2}\right], \quad l(\Delta) \text{ is even},
\end{equation}
\begin{equation}\label{29}
\dim{\mathcal{A}l^o_n}=p_o(n)-\left[\dfrac{n+1}{2}\right], \quad l(\Delta) \text{ is odd},
\end{equation}
where $p_{e/o}(n)$ is the number of partitions of $n$ into even/odd number of integers.
\end{statement}
\end{framed}

Let us apply this claim to the Alexander polynomial itself. As it was discussed, the Alexander polynomial has a symmetry with respect to the transposition of the diagram \eqref{ranklevel}. In terms of the loop expansion, it implies that
\begin{equation}
\sum_{|\Delta|=n}\alpha_\Delta C_\Delta (R^T) = (-1)^n \sum_{|\Delta|=n} (-1)^{l(\Delta)} \alpha_\Delta C_\Delta (R).
\end{equation}
At the same time, we established that solutions to the Alexander system can be either even or odd with respect to this operation, depending on the length of the partitions contained in the expression. Hence, the Alexander polynomial consisting of these solutions, satisfies
\begin{statement}\label{s2}
The group factors in the expansion of the Alexander polynomial have the following structure:\\
in every order $n$ of $\h$, the terms of degree $k= n \ {\rm mod} \ 2$  consist of even monomials $C_\Delta$, and the terms of degree $k= n+1 \ {\rm mod} \ 2$  consist of odd monomials.
\end{statement}

We also state that every odd solution $X^o_{n,m}(C)$ of the Alexander system can be expressed in a polynomial combination of even solutions $X^e_{n,m}(C)$. For example,
\begin{eqnarray}
X^o_{5,1}(C) &=& C_1 X^e_{4,1}(C) \nonumber \\
X^o_{6,1}(C) &=& \frac{1}{3}C_2 X^e_{4,1}(C) - \frac{1}{3}C_1 X^e_{5,1}(C)  \\
X^o_{6,2}(C) &=& C_2 X^e_{4,1}(C) \nonumber
\end{eqnarray}

\section{The KP hierarchy}
Now we need a few standard results about the KP hierarchy \cite{MiwaJimboDate,Kharchev}.
\\

The defining property of this system is the Hirota bilinear identity satisfied by the KP $\tau$-function. It can be written in a simple form using the Hirota derivatives $D_x$ defined as follows:
\begin{equation}
f(x+y)g(x-y)=\sum^{\infty}_{j=0}\frac{1}{j!}(D^j_x f\cdot g)y^j.
\end{equation}
The case of multiple variables is defined in a similar manner via the multivariate Taylor expansion. Then, the Hirota bilinear identity is written as
\begin{equation}
\oint \dfrac{d z}{2\pi i} e^{ 2 \sum _i z^i t_i } e^{\sum _i -\frac{1}{i z^i}D_{T_i}}e^{\sum _j t_j D_{T_j}} \tau \otimes \tau  = 0,
\end{equation}
where  $D_{T_i}$ are the Hirota derivatives, $\tilde D \equiv  (D_{T_{^1}}$, ${1\over 2} D_{T_{^2}},\ldots$), and $x=\{ x_1,x_2,\dots \}$ are formal parameters.
Consequently, it can be shown that  the KP hierarchy is defined by the following generating function:
\begin{equation}\label{hir}
\sum ^\infty _{i=0} \chi_i(-2t)\chi_{i+1}(\tilde D_T) e^{[\sum _jt_jD_{T_j}]} \tau \otimes \tau  = 0,
\end{equation}
where  $\chi_i$ are the Schur functions in symmetric representations. The KP equations in the Hirota form appear as coefficients of expansion of (\ref{hir}) in $t_i$ in a polynomial form $P_n(D_1,D_2,\ldots,D_{n-1})$. The first few equations are: \\
$\bullet \quad$ The equation of order 4, which is the KP equation itself:
\begin{flushleft}
\begin{equation}
[4D_1 D_2 - 3D_2^2-D_1^4]\tau \otimes \tau = 0
\end{equation}
\end{flushleft}
$\bullet \quad$ Higher KP equations:
\begin{align}\label{Hirota5}
P_5 \  & =3D_1 D_4 - 2D_2 D_3 - D_1^3 D_2 \nonumber; \\[4pt]
P_{6,1}& =D_1^2 \left(D_1^4-4 D_3 D_1+3
   D_2^2\right),\nonumber \\
P_{6,2} & =D_1^6-20 D_3 D_1^3-45 D_2^2
   D_1^2+144 D_5 D_1-80 D_3^2,\nonumber \\
P_{6,3} & =D_1^6+10 D_3 D_1^3-36 D_5 D_1-20 D_3^2+45 D_2
   D_4;   \\[4pt]
 P_{7,1} & =   D_1 D_2 \left(D_1^4-4 D_3 D_1+3
   D_2^2\right),\nonumber \\
P_{7,2} & =   D_2 D_1^5-5 D_4 D_1^3+50 D_2 D_3 D_1^2-10 D_2^3
   D_1-80 D_6 D_1+20 D_3 D_4+24 D_2 D_5 , \nonumber\\
P_{7,3} & =   D_2 D_1^5-5 D_4 D_1^3+20 D_2 			D_3 D_1^2-10 D_2^3 D_1-20
   D_6 D_1-10 D_3 D_4+24 D_2 D_5,\nonumber \\
P_{7,4} & =D_2
   D_1^5+10 D_4 D_1^3-40 D_2 D_3 D_1^2+5 D_2^3 D_1+40 D_6
   D_1-40 D_3 D_4+24 D_2 D_5 . \nonumber
\end{align}

An important property of the Hirota derivative is that an odd number of such operators acts trivially. We are interested in the space of differential polynomials that form the KP hierarchy, therefore we do not want to distinguish between two polynomials differing by an odd monomial. Hence, we are looking for even polynomials by symmetrizing everything w.r.t. $D_k \rightarrow -D_k$.\\

Due to \cite{TG}, the generating function \eqref{hir} can be expanded using the Weyl character formula. The coefficient of $\chi_Y(t)$ gives rise to the equation
\begin{equation}
\begin{vmatrix}
\chi_{f_1+1} (-\tilde{D}/2) & \chi_{f_1+1} (\tilde{D}/2) &\ldots & \chi_{f_1+m-1} (\tilde{D}/2) \\
\chi_{f_2} (-\tilde{D}/2) & \chi_{f_2} (\tilde{D}/2) &  \ldots & \chi_{f_2+m-2} (\tilde{D}/2) \\
\vdots & \vdots & \vdots \\
\chi_{f_m-m+2} (-\tilde{D}/2) & \chi_{f_m-m+2} (\tilde{D}/2) &  \ldots & \chi_{f_m} (\tilde{D}/2) \\
\end{vmatrix} \tau \otimes \tau =0,
\end{equation}
where $f_1 \geq f_2 \geq \ldots \geq f_m \geq 1$, $m \geq 2 $ and $Y=[f_1,f_2,\ldots f_m]$. The graded order of such an equation is $f_1+f_2+\ldots f_m +1$. \\
In \cite{TG}, it has been proven that all of these polynomials are independent. However,  we are interested in the symmetrized version and apparently some of these polynomials only differ by odd polynomials. \\
There is a more suitable expression, however.
By \cite{Natanzon}, the differential basis of the KP hierarchy is given by polynomials numbered by the two row  diagrams, $Y=[i,j], \ i\geq j \geq 2$:
\begin{equation}
\begin{vmatrix}
\chi_{i} (-\tilde{D}/2) & \chi_{i} (\tilde{D}/2) & \chi_{i+1} (\tilde{D}/2) \\
\chi_{j-1} (-\tilde{D}/2) & \chi_{j-1} (\tilde{D}/2) & \chi_{j} (\tilde{D}/2) \\
1 & 1 & \dfrac{1}{2}D_1
\end{vmatrix} \tau \otimes \tau =0,
\end{equation}
This means all other KP equations are obtained by multiplying these basis equations by a suitable monomial $D_\Delta$. We see here that this differential basis is naturally numbered by two row partitions with at least 2 hooks. We have two vector spaces here. One is formed by these basis KP equations, which we denote by $\KP_n$. Its dimension is easy to compute. The other one is the space of \emph{all} possible polynomial differential operators that annihilate the $\tau$-function. It is denoted by $\fullKP_n$ and generated multiplicatively by $\KP_n$, but we also want to look at it as a vector space.

\section{The main results}

As we reminded in the previous section, the generation function of the  KP hierarchy is the integral bilinear identity

\begin{equation}
\oint \dfrac{d z}{2\pi i} e^{2 \sum _i z^i t_i } e^{-\sum _i \frac{1}{i z^i}D_{T_i}}e^{\sum _i t_i D_{T_i}} \tau \otimes \tau  = 0.
\end{equation}

Now replace the Hirota operators with the Casimir eigenvalues and symmetrize the identity w.r.t. $C_i \leftrightarrow -C_i $ to guarantee that the odd number of Hirota operators cancels $\tau \otimes \tau$:\\
\begin{equation}\label{a}
B:=\oint \dfrac{d z}{2\pi i} e^{ 2 \sum _i z^i t_i } e^{\sum _i -\frac{1}{i z^i}C_i}e^{\sum _i t_i C_i}  +
\oint \dfrac{d z}{2\pi i} e^{ 2 \sum _i z^i t_i } e^{ \sum _i \frac{1}{i z^i}C_i}e^{- \sum _i t_i C_i} .
\end{equation}
Our main claim is that vanishing of $B$ is equivalent to the Alexander property (\ref{a-a}):
\begin{equation}
\boxed{
\ B=0 \ \ \Longleftrightarrow \ \ \A^\K_R(q)=\A^\K_{[1]}(q^{\vert R\vert}). \
}
\end{equation}

Let us put it differently. One may notice that some Hirota equations \eqref{Hirota5} are identical to solutions of the Alexander system \eqref{alexsol}. Moreover, by computer check, we found that the spaces $\Al^e_n$ and $\fullKP_n$ literally coincide when we substitute $C_k \leftrightarrow D_k$, at least,  up to order 17.
The table compares the dimensions of these spaces up to order 17:
\begin{table}[H]
\begin{center}
\begin{tabular}{c|c|c|c}
n & dim $\mathcal{A}_n$ & dim $\mathcal{KP}_n$ & dim $\widehat{\mathcal{KP}}_n$\\
\hline
4&1&1&1\\
5&1&1&1\\
6&3&2&3\\
7&4&2&4\\
8&8&3&8\\
9&10&3&10\\
10&17&4&17\\
11&22&4&22\\
12&34&5&34\\
13&43&5&43\\
14&62&6&62\\
15&79&6&79\\
16&110&7&110\\
17&138&7&138\\
\end{tabular}

\end{center}
\end{table}
In these terms, our main result is proving that $\fullKP_n$ and $\Al^e_n$ are literally the same polynomial spaces. This means that the group factors, which appear from the basic property (\ref{a-a}) of the Alexander polynomial, mysteriously correspond to equations of the KP hierarchy.  This statement also gives an expression for the number of actually different KP equations of each order: formulas (\ref{28})-(\ref{29}), which, to our best knowledge, has not been known so far. This is another our result in this paper.

The whole proof consists of proving that the solutions of the Alexander property satisfy the bilinear identity, and then proving that the differential KP basis actually generates the space $\Al^e_n$. We leave the complete proof for a separate paper, and restrict ourselves here to prove that {\bf the Hirota KP bilinear equations $B$ are solved when the Hirota derivatives are replaced by the Casimir eigenvalues in the 1-hook diagrams, which also solves the Alexander property (\ref{a-a}) or (\ref{16})}.

Indeed, the hook value of the Casimir invariants are given by \eqref{casrl}. For simplicity denote $x= r-1/2, y=-L-1/2$, then the exponential in the integrand  becomes
\begin{equation}
e^{-\sum _i \frac{1}{i z^i}C_i}=e^{-\sum _i \frac{1}{i z^i}(x)^i}e^{\sum _i \frac{1}{i z^i}(y)^i}=\dfrac{z-x}{z-y},
\end{equation}
while the same exponential in the second integral is just the same to the power $-1$.\\
This allows us to compute the integrals taking the residues at $z_2=x, \ z_1=y$. Therefore, restricted to 1-hook diagrams, \eqref{a} vanishes:
\begin{equation}
(y-x)e^{-\sum t_i(x^i+y^i)}+(x-y)e^{-\sum t_i (x^i+y^i)}=0.
\end{equation}
We see that the Hirota equations vanish when the Hirota derivatives are replaced by the Casimir eigenvalues for the 1-hook diagrams.

\section{Discussion}

Our observation immediately puts a set of  questions.

The first question is about the relation of the KP hierarchy with the HOMFLY invariants. At the moment, it is only known a relation of these with integrable systems in the case of torus knots and links \cite{inttorus}. In this paper, we also see
that somehow two "opposite" limits of these polynomials {\it for arbitrary knots} encode a lot of data about the hierarchy. It would be interesting to see what more we could derive from the HOMFLY invariant for an arbitrary knot/link.

Second, we see that the equations of the KP hierarchy are related to 1-hook diagrams, which do not arise in the explicit definition of the equations. Therefore, we get some new combinatorial property of the Hirota equations of the KP hierarchy. It is very interesting to understand its nature.

\section*{Acknowledgements}


This work was funded by the Russian Science Foundation (Grant No.16-12-10344).

\end{document}